\documentclass[a4,11pt]{article}
\usepackage{epsfig}
\usepackage{amssymb,epsfig,graphicx,epsf,epstopdf}

\usepackage[titletoc,toc,title]{appendix}

\usepackage{amsfonts}
\usepackage{amsmath}
\usepackage{feynmf}

\usepackage{latexsym}
\usepackage{float} %Posiciona la imagen donde la pones el codigo exactamente.

\usepackage{slashed}
\usepackage{cite}
\usepackage[utf8]{inputenc} % acentos! Tildes en realidad
\usepackage{mathtools}
\usepackage[font=small,skip=-28pt]{caption}
\usepackage{color}
\DeclareFontFamily{OT1}{pzc}{}
\DeclareFontShape{OT1}{pzc}{m}{it}{<-> s * [1.10] pzcmi7t}{}
\DeclareMathAlphabet{\mathpzc}{OT1}{pzc}{m}{it}

\usepackage{commath} % Para módulo como \abs{}
\usepackage{accents} % Punto para la derivada \dt{} o \ddt{}
\usepackage{relsize} %agranda texto adentro de una ec. {\mathlarger{(x+y=z)}}
\usepackage[makeroom]{cancel} % Para tachar algo en una ecuacion
\usepackage{xcolor}
\usepackage[colorlinks=true, allcolors=black]{hyperref} % Click in reference and goes there.

\newcommand{\ben}{\begin{equation}}
	\newcommand{\een}{\end{equation}}
\newcommand{\be}{\begin{equation*}}	
	\newcommand{\ee}{\end{equation*}}

\newcommand{\ba}{\begin{eqnarray}}
\newcommand{\ea}{\end{eqnarray}}
\newcommand{\bal}{\begin{aligned}}
\newcommand{\eal}{\end{aligned}}

\begin{document}

\title{Condensates beyond the horizons}

\author{Jorge Alfaro\\
Pontificia Universidad Cat\'olica de Chile, Av. Vicu\~na Mackenna 4860,\\ Santiago, Chile.\\ 
\\
Dom\`enec Espriu and Luciano Gabbanelli\\
Departament of Quantum Physics and Astrophysics and\\ 
Institut de Ci\`encies del Cosmos (ICCUB), Universitat de Barcelona,\\
Mart\'\i ~i Franqu\`es 1, 08028 Barcelona, Spain.}

\date{}

\maketitle

\begin{abstract}
In this work we continue our previous studies concerning the possibility of the existence of a Bose-Einstein condensate in the interior of a static black hole, a possibility first advocated by Dvali and G\'omez. We find that the phenomenon seems to be rather generic and it is associated to the presence of an horizon, acting as a confining potential. We extend the previous considerations to a Reissner-Nordstr\"om black hole and to the de Sitter cosmological horizon. In the latter case the use of static coordinates is essential to understand the physical picture. In order to see whether a BEC is preferred, we use the Brown-York quasilocal energy, finding that a condensate is energetically favourable in all cases in the classically forbidden region. The Brown-York quasilocal energy also allows us to derive a quasilocal potential, whose consequences we explore. Assuming the validity of this quasilocal potential allows us to suggest a possible mechanism to generate a graviton condensate in black holes. However, this mechanism appears not to be feasible in order to generate a quantum condensate behind the cosmological de Sitter horizon.

\end{abstract}
\vfill
\noindent

\noindent
ICCUB-19-006

\newpage

\section{Introduction}
Some time ago Dvali, G\'omez and coworkers\cite{DG1}
suggested that black holes (BH) could be understood as
Bose-Einstein condensates of gravitons. This provocative idea could lead to an 
alternative understanding of some of the most striking features of BH. For instance, it was suggested that
Hawking radiation \cite{Radiation} would be due to the leakage from the condensate. 
This picture could bring new ideas about the Bekenstein entropy \cite{Entropy}, the absence of 
hair \cite{NoHair} and the perennial discussion on loss of information in BHs \cite{InformationLoss} too.

In \cite{AEG} we elaborated on some of these ideas, deviating somehow from the
original line of thought of the authors (see \cite{Germani} for a different approach). We assumed the pre-existence of a classical gravitational 
field created by an unspecified source that generates the Schwarzschild metric. Nothing can classically 
escape from the interior of a BH, so in practice this situation would correspond to a confining potential 
where quantum fields are presumably trapped, gravitons in particular.

Continuing with this analogy, these trapped gravitons as well as other quanta present
have had a long time to thermalize in a static Schwarzschild BH and it is natural 
to expect that eventually a  Bose-Einstein condensate (BEC) could form.
Of course these `gravitons' are in no way freely propagating transverse gravitons. 
They are necessarily off-shell ($q^2\neq 0$) and thus have some sort of effective mass. 

In \cite{AEG} we tried to substantiate this picture and propose a  consistent set 
of equations describing a BEC constructed 
on top of the classical field created by a BH, that could be identified as the counterpart of the 
Gross-Pitaevskii equation\cite{GPEquation}. We did succeed and, remarkably enough, the characteristics 
of the resulting BEC are uniquely described in terms of the Schwarzschild radius of the BH and the value 
of a dimensionless parameter, interpreted as a chemical potential. 
A condensate appears not only to be possible but actually intimately related to the classical field that 
sustains it and determines its characteristics. 

In the present study, after reviewing the main characteristics of this proposal in the context of a Schwarzschild BH, we
also consider other geometries that exhibit an horizon, such as Reissner-Nordstr\"om or de Sitter. We find that BEC-like
solutions may be present in all these cases too. We then proceed to study the energy associated to these
solutions to see whether they are energetically favourable. As it is well known, the definition of energy is problematic
in general relativity. Here we approach this issue in the light of a quasilocal definition of energy, whose
properties and main characteristics, in particular in the context of a possible BEC, are presented in section \ref{QLE}.
The quasilocal energy is used
to define a quasilocal potential that in turn is applied to attempt to understand the binding of matter falling
in the black hole. Associated to this, we also provide a tentative explanation of the origin of the `graviton' condensate itself. 
In sections \ref{RRNN} and \ref{dS} we extend the analysis of the quasilocal energy to the Reissner-Nordstr\"om and de
Sitter solutions.

\section{Summary of the Schwarzschild results}
The main point of the work of Dvali and G\'omez is that the physics of BH can be understood 
in terms of a single number $N$, the number of (off-shell) gravitons contained in the Bose-Einstein condensate (BEC).
These condensed gravitons have a wavelength $\lambda\sim \sqrt N L_P$, $L_P$ being the Planck length. They have 
a characteristic interaction strength $\alpha_g\sim 1/N$. The mass 
of the BH is $M\sim \sqrt N M_P$ and its Schwarzschild radius therefore is given by
$r_s\sim \sqrt N  L_P$, thus agreeing with the Compton wavelength of the quantum  gravitons, in
accordance with the uncertainty principle that dictates $\lambda \simeq r_s$ in the ground state of
the quantum system. Therefore, up to various factors of the Planck mass everything is governed by $N$, the
number of intervening gravitons. Modulo some assumptions, all these results stem 
from the basic relation postulated in \cite{DG1} (unless otherwise stated we work in units where $c=\hbar=1$)
\begin{equation}
	r_g= \frac{N}{M}\end{equation}
that relates the number of gravitons $N$, the mass of the gravitating object $M$ and its gravitational
radius $r_g$. For a Schwarzschild BH $r_g=r_s$.
In \cite{AEG} the previous relations were rederived from rather simple assumptions.

Bose–Einstein condensation of a spatially homogeneous gas with attractive 
interactions is normally precluded by the existence of a conventional phase transition into either 
a liquid or solid. Even when such a condensate can form, its occupation 
number is low. Repulsive forces act to stabilize the condensate against collapse, but gravitons do not have 
repulsive interactions, at least naively, and therefore one would conclude that a BEC is impossible to 
sustain, particularly as we expect $N$ to be very large. However, it is up to the equations themselves to establish
whether such a condensate is possible or not. In addition, in theories of emergent gravity, the ultimate 
nature of gravitons may involve some kind of fermionic degrees of freedom (e.g. in the model suggested in \cite{AEP}). 
Then, repulsion is assured at some scale and fundamental collapse prevented. In fact, the results in \cite{AEG}
suggest that condensation is possible.

Let us now summarize our findings as reported in \cite{AEG}. A Gross-Pitaevskii-like equation is obtained in the following way: the graviton condensate would necessarily be described by a tensor field $h_{\alpha\beta}$ representing dimensionless fluctuations of a classical background metric $\tilde g$. We restricted ourselves to perturbations that respected the spherical symmetry (and thus with zero angular momentum), describing excitations of the metric. Although $h_{\alpha\beta}$ is inspired by a perturbation to a background metric, in this work it is considered an independent field and therefore treated to all orders in perturbation theory. $h_{\alpha\beta}$ is not necessarily ``small".

In the picture of \cite{DG1} if a BEC is present with `gravitons' acquiring all the same momenta $\sim 1/r_s$ and being weakly interacting for a macroscopic BH (recall $\alpha_g\sim 1/N$) the total energy stored in the condensate should be $\sim N/r_s$ and this quantity would be a conserved one. Therefore, $N$ would be conserved too. The previous reasoning shows very clearly that the `gravitons' contemplated in the present scenario (`maximum packing'), if realized, have nothing to do with freely propagating gravitons as the number of massless on-shell gravitons is not a conserved quantity. 

The energy of the gravitons contained in a given volume would be
\begin{equation}
	E=\frac{M_P{}^2}2 \int dV \varepsilon^2\ h_{\alpha\beta}\ h^{\alpha\beta}= \int dV \varepsilon \,\rho_{h}\ ,\end{equation}
where we assume that the energy per graviton $\varepsilon$ is constant and approximately given by $\varepsilon= 1/\lambda$ with $\lambda=r_s$, and
\begin{equation}\label{Density}
	\rho_{\hat h}\equiv \frac12\ \frac{M_P{}^2}{\lambda}\, h_{\alpha\beta}\, h^{\alpha\beta}\ ,\end{equation}
While there is no formally conserved current, the above quantity can be interpreted as a `graviton' number density, and the integral of the graviton density (\ref{Density}) in the interior of the BH has to be interpreted as the number of constituents of the `graviton' BEC.

The chemical potential term in the action should be conjugated to the graviton density of the condensate inside a differential volume element $dV$. In order to respect diffeomorphism invariance, the simplest form of introducing such a term is by means of
\begin{equation}\label{ActionCPTerm}
	\Delta S_{chem. pot.}
	= -\tfrac{1}2\, M_P{}^2\int d^4 x\sqrt{-\tilde g}\,\mu \, h_{\alpha\beta} h^{\alpha\beta}\ .\end{equation} 
This term does resemble a mass term for the spin-2 excitation and indeed it is some sort of effective mass in practice as the `gravitons' in the BEC are supposed to be quasi-non-interacting in the large $N$ limit \cite{DG1}. However, GR eventually requires for the quantity $\mu$ to actually be position dependent,\footnote{This is easily seen by requiring the covariant conservation of the equations of motion. See \cite{AEG}.} i.e. $\mu=\mu(r)$, and transform as a scalar.

The total action would then be
\begin{equation}\label{ActionFull}
	S(h)={M_P}^2\int d^4x\sqrt{-g}\,R(g)-\tfrac12\,{M_P}^2\int d^4x \sqrt{-\tilde g}\,\mu\,h_{\alpha\beta}h^{\alpha\beta}\ .\end{equation}
Indices are raised and lowered using the full metric $g_{\alpha\beta}=\tilde g_{\alpha\beta}+h_{\alpha\beta}$ in order to preserve diffeomorphism invariance that is an exact invariance of the above action.
In fact, part of this statement is shown in \cite{Weinberg}: under an infinitesimal displacement in the coordinates given by $\delta_{\scriptscriptstyle D} x^\mu=-\xi^\mu(x)$, the full metric changes as $\delta g_{\mu\nu}=\xi^{\rho}\ g_{\mu \nu, \rho} + \xi^{\beta}_{\ ,\nu}\ g_{\mu \beta} + \xi^{\beta}_{\ ,\mu}\ g_{\nu \beta}$. The same transformation applies to the background metric $\tilde g_{\mu\nu}$. As the perturbation is defined by the following, $h_{\mu\nu}= g_{\mu\nu}-\tilde g_{\mu\nu}$, under the same perturbation of the coordinate system, the same rule of covariant transformation is obtained.  Together with the fact that the chemical potential $\mu$ behaves as a scalar under a general coordinate transformation, ensure automatically the general covariance of the theory.
Note that the determinant of the background metric is used in the measure of the chemical potential term. Starting with (\ref{ActionFull}), a fully covariant expansion in powers of $h_{\alpha\beta}$ can be performed up to the desired order of accuracy. 
By construction these equations will of course be non-linear (like Gross-Pitaevskii's), the LHS is just the `Schrödinger' equation (like Gross-Pitaevskii's) and it contains an additional piece proportional to the chemical potential (like Gross-Pitaevskii's too). 
It is worth noting again that the action (\ref{ActionFull}) is invariant under general coordinate transformations, but it is not background independent and it should not be. 

The corresponding equations of motion\footnote{This is slightly modified with respect to the proposal in \cite{AEG} as
in the latter an inconsistency was detected regarding the angular coordinates. This does not affect the essence
of the discussion. The present equation of motion and action are fully consistent.} 
\begin{equation}\label{EEq}
	G_{\alpha\beta}(\tilde g+h)=\mu\,\sqrt{\tfrac{\tilde g}{g}}\, \bigl(h_{\alpha\beta}-h_{\alpha\sigma}h^\sigma{}_\beta\bigr)\,,\end{equation}
are interpreted as the Gross-Pitaevskii equations describing the properties of a graviton condensate sitting in the BH interior. 
Retaining the leading order in $h_{\alpha\beta}$ on the RHS 
(\ref{EEq}) takes a simpler form, very similar to the familiar Gross-Pitaevskii equation 
\begin{equation}
	G_{\alpha\beta}(\tilde g+h)= \mu\, h_{\alpha\beta}\end{equation}
that is good enough to draw the main physical conclusions qualitatively. We shall however adhere to (\ref{EEq})
that can be derived from an action principle.

Of course it could be that the only consistent solution of (\ref{EEq}) would correspond to
$\mu=0$, but this is not the case. While it can be seen that under very reasonable hypothesis
the only solution for $r>r_s$ is $\mu=0$ and $h_{\alpha\beta}=0$, in the interior of the BH we found in \cite{AEG} a family of solutions that could be interpreted as a condensate structure.
These solutions take the form
\begin{equation}\label{hTensor}
	h_{\alpha\beta}=\text{diag}\left(\frac{\varphi}{1-\varphi}\,\tilde g_{tt}\,;\,\frac{\varphi}{1-\varphi}\,\tilde g_{rr}\,;\,0\,;\,0\right),\end{equation}
where $\varphi$ is $r-$independent.
This implies that the total metric is
\begin{equation}\label{MetricFullGeneric}
	g_{\alpha\beta}=\text{diag}\left(\frac{1}{1-\varphi}\,\tilde g_{tt}\,;\,\frac{1}{1-\varphi}\,\tilde g_{rr}\,;\,\tilde g_{\theta\theta}\,;\,\tilde g_{\phi\phi}\right). \end{equation}
The parameter  $\varphi$ satisfies the following (Gross-Pitaevskii -like) equation\footnote{Note that the following equation contains no derivative terms, like e.g.
  the corresponding Gross-Pitaevskii equation describing a uniform gas of interacting atoms: $\varphi^2= \mu/g$, where $g$ would be the coupling constant and $\varphi$
  the BEC order parameter.}
\begin{equation}\label{GPEquation}
	-\frac{\varphi}{r^2}=\mu\left(1-\varphi\right)^2\varphi\ .\end{equation} 
with a constant $\varphi$ and where the chemical potential is
\begin{equation}\label{ChemPot}
	\mu=\frac{\mu_0}{r^2}\ , \qquad \mu_0=-\frac{1}{(1-\varphi)^2}\ . \end{equation}
Assuming the `maximum packaging' condition $\lambda= r_s$ \cite{DG1}, according to our previous discussion, 
the number of particles would be given by the integral 
\begin{equation}\label{Integralh2}
	\int_0^{\infty} d^3x \sqrt{-\tilde g}\ {h_\alpha}^\beta{h_\beta}^\alpha =4\pi\int_0^{r_s} dr\, r^2\ ( h_t{}^{t\,2} + h_r{}^{r\,2}) = \frac{8\pi}{3}\,{r_s}^3\,\varphi^2\ ,\end{equation}
that states that the integral of the square modulus of the wave function has a constant value. The volume element 
for the background metric $d^3x\sqrt{-\tilde g}$ has been used. Then
\begin{equation}\label{mistery}
	N=\frac{4\pi}{3}\,{M_P}^2\, \varphi^2 \,{r_s}^2 \qquad \Longrightarrow\qquad r_s= \sqrt{\frac{3N}{4\pi\varphi^2}}\ L_P\ , \quad 0\le \varphi < 1 \ .\end{equation}
Here again the upper limit for the wave function enters explicitly; if $\varphi\rightarrow1$, the metric becomes singular. These relation agrees nicely with the proposal of Dvali and G\'omez. The rest of relations of their work can basically be derived from this. 

Possibly our more striking results are that the dimensionless chemical potential $\mu_0= \mu(r) r^2$ stays constant and 
non-zero throughout the interior of the BH, and that so does the quantity $h_\alpha\,^\alpha=\varphi$ previously 
defined and entirely determined by the value of $\mu_0$, and viceversa. It is natural to interpret $\mu_0$ as the variable
conjugate to $N$, the number of gravitons.

Of course the solution $\varphi=0$ is also possible. If $\mu_0=0$ the only solution is $\varphi=0$, $N=0$ reproducing the familiar Schwarzschild metric in the BH interior. However, the interesting point is that solutions 
with $\varphi\neq 0$ exist and the
value of $\mu_0$ is uniquely given by \eqref{ChemPot}. Assuming continuity of the solution the limit $\varphi\to 0$ 
would correspond to $\mu_0=-1$. This dimensionless number would then be a unique property of a Schwarzschild BH.

$\mu_0$ has a rather peculiar behaviour. It is non-zero inside and at the event horizon. Outside it appears to be exactly zero.
If we forget about the geometrical interpretation of BH physics and let us treat the problem as a 
collective many body phenomenon,  it is clear that `gravitons' are trapped behind the horizon: the jump 
of the chemical potential at $r=r_s$ would prevent 
the `particles' inside from reaching infinity. From this point of view it is quite natural to have a lower chemical potential
inside the horizon than outside (where is obviously zero) as otherwise the configuration would be thermodynamically unstable.
In the present solution particles (`gravitons' in our case) cannot escape.

It is to be expected that other quanta may form condensates too, being described by Gross-Pitaevskii-like equations
similar in spirit to the one considered for gravitons. This would lead to the introduction of chemical potentials conjugate to their respective number of particles. However it is not clear a priori which equations they obey;
in particular which potential energy one should use in these cases. We will advocate here to use a potential derived from
the quasilocal energy and investigate some possible consequences of this.

\section{Quasilocal energy}\label{QLE}
It is a fundamental tenet of general relativity that there is no such concept as a local energy of the gravitational field.
Ambitious efforts to associate energy to extended, but finite, spacetime domains have resulted in 
quasilocal definitions that could associate the configurations of the gravitational field to an energy principle.
Although several approaches have led to some interesting results, no definitive consensus for the expression of the
quasilocal energy seems to have emerged (various proposals for quasilocal energies are 
reviewed in \cite{Szabados} and references therein).

In what follows we will use the quasilocal energy elaborated by  Brown and York\cite{BrownYorkMath} in order to analyze 
its behaviour for several spacetime configurations where horizons are present, with or without the
possible graviton condensates just discussed. This magnitude is one of the promising candidates for obtaining the quasilocal energy in general relativity.
According to this proposal, a suitable fixing of the metric over the boundaries \cite{YorkKAction} provides a well-defined
action principle for gravity and matter, and dictates a natural choice for the definition of the quasilocal energy contained in the region of interest without the need for any other geometric structures. 
Besides, its properties can be obtained through a Hamilton-Jacobi type of analysis involving the canonical action.
Among these properties we could mention that the proposal of
\cite{BrownYorkMath} agrees with the Newtonian limit for spherical stars, it is applicable to thermodynamic
problems\cite{Braden,York} and the asymptotic limit is the Arnowitt-Deser-Misner \cite{ADM} expression for
the energy at infinity for asymptotically flat spacetimes \cite{Regge}. Physically, the Brown and York proposal
is meant to represent the total energy enclosed by a space-like surface. Given that the idea of quasilocal energy may not be familiar to several readers,
we provide in the Appendix \ref{quasilocalenergy} a short explanation.

Apart from the intrinsic interest of defining an energy, using a least-energy principle could allow us to determine whether 
a `graviton' condensate is in some sense preferred to the standard no-condensate situation. One might after all expect that
matter falling into a black hole will eventually thermalize and the formation of a BEC may not seem such an exotic 
possibility then, but in the case of gravitons the formation of the condensate is a lot less intuitive. Minimization of 
the quasilocal energy may be a convenient tool to answer whether a BEC for gravitons exists or not.

Let us review the definition of the quasilocal energy derived in \cite{BrownYork}. The basic idea is to express the energy as a variation of the action with respect to fixed endpoints.
We will deal with the physics of a compact spacetime domain $M$ that can be topologically decomposed as the product of a spatial three-space $\Sigma$ times a real line interval. At any time, each compact hypersurface $\Sigma$ is bounded by a two-surface denoted as $\partial\Sigma=B$. 
In our case, $\Sigma$ is the interior of a $t=$ constant slice with a two$-$boundary $B$ given by spheres concentric with the Schwarzschild black hole determined by $r=$ constant. The quasilocal energy in $\Sigma$ will be given by the following definition
\ben\label{QLEnergy}
	E=\frac{1}{8\pi G} \int_{B} d^2x \sqrt{\sigma}\left(K-K_0\right) \, \een
where $\sigma$ is the determinant of the induced two-metric $\sigma_{ij}$ on the boundary $B$; $K$ is the trace of the extrinsic curvature $K_{ij}$ of this two-boundary, embedded into the space-like hypersurface $\Sigma$; and $K_0$ corresponds to the energy of the same two-boundary but embedded in a Minkowskian vacuum. There is a freedom in choosing a ground level of energy, hence the latter term acts as a reference term that must be subtracted to obtain the physical energy.

Notice that the separation of the four dimensional space in a product $\Sigma\times [t',t'']$ implies that the quasilocal energy is invariant under general coordinate transformations on $\Sigma$ but not under all four dimensional general coordinate transformations, which also involve time. Therefore quasilocal energy will be dependent on the proper time of the observer \cite{WuChenLiu}.

In the present study we limit ourselves to static and spherically symmetric metrics of the form
\begin{equation}\label{Metric}
	ds^2=-\epsilon\,N(r)^2\,dt^2+\epsilon\,f(r)^{-2}\,dr^2+r^2\,d\Omega^2\ ;\end{equation}
with such definition $N(r)$ and $f(r)$ are always positive functions of the coordinate $r$ and $\epsilon$ is either $1$ or $-1$. 
If we are placed outside the horizon, $t$ is a timelike coordinate, then the temporal metric potential $g_{tt}$ is negative 
and in this region $\epsilon=1$. At the horizon $g_{tt}=-\epsilon\, N^2$ and $g_{rr}=\epsilon\,f^{-2}$ exchange sign, 
hence in the interior $\epsilon=-1$. A spacelike slice has the following associated metric
\begin{equation}\label{MetricSigmaSphSym}
	\gamma_{ij}\,dx^idx^j=\epsilon\,f^{-2}\,dr^2+r^2\,d\Omega^2\ ;\end{equation}
and the unit normal vector to constant $r$ surfaces are chosen to be future-pointing
\ben\label{Normal}
	n^i=\epsilon\,f(r)\,\delta_r^i \ .\een
This way of defining objects, factorizing the signs via $\epsilon$, makes the extension of the definition of the 
quasilocal energy to the interior zone, beyond the event horizon, straightforward as pointed in \cite{Lundgren}, where it is argued on physical grounds why the normal vector to constant $r$ surfaces has to change sign at $r=r_s$. We shall adhere to this prescription.

The extrinsic curvature of the two-boundary $B$ as embedded in $\Sigma$ is defined by the covariant derivative in the 
spacelike slice of the normalized tangent vector to this hypersurface $\Sigma$
\begin{equation}\label{ExtrinsicCurvatureB}
	K_{ij}=-\sigma_i^k\,\nabla_k\,n_j \end{equation}
along a direction projected by the operator
\begin{equation}
	\sigma_{ij}=\gamma_{ij}-n_in_j\ .\end{equation}
For the calculation it is only needed the following connection coefficients
\begin{equation}
	\Gamma_{rr}^r=-\frac{f'}{f}\ ; \hspace{80pt} \Gamma_{r\theta}^\theta=\Gamma_{r\phi}^\phi=\frac{1}{r}\ ; \end{equation}
the prime denotes $r$ derivatives. The trace of the extrinsic curvature needed for computing the quasilocal energy is
\begin{equation}
	K=-2\,\epsilon\,\frac{f(r)}{r}\end{equation}
and the subtraction term $K_0$ is obtained embedding the sphere $B$ in a Minkowskian spacetime by setting $f(r)=\epsilon=1$.
The next step is straightforward, putting everything together in \eqref{QLEnergy} with the determinant of the induced 
two-metric given by $\sqrt{\sigma}=r^2\sin^2\theta$, the quasilocal energy is 
\begin{equation}\label{QLEnergyIntegrated}
	E=\frac{r}{G}\left[1-\epsilon\,f(r)\right]\ . \end{equation}

\subsection{Quasilocal energy in a Schwarzschild black hole}
Let us now consider the change in the quasilocal energy, as defined above, when a condensate with spherical symmetry,
such as the one previously described, exists in a static and spherically symmetric BH.
The significant metric elements in \eqref{Metric}, both within and beyond the event horizon, are obtained
particularizing \eqref{MetricFullGeneric} to the Schwarzschild solution
\begin{equation}\label{MetricSchw}
  N(r)=\sqrt{\frac{\epsilon}{1-\varphi}\left(1-{\frac{r_s}{r}}\right)}\ ; \hspace{30pt}
  f(r)=\sqrt{\epsilon\left(1-\varphi\right)\left(1-{\frac{r_s}{r}}\right)}\ .\end{equation}
In the outer region we get the known solution by setting $\epsilon=1$ and $\varphi=0$. In the interior $r$ is a timelike coordinate,
$\epsilon=-1$ and $\varphi$ may be different from zero. The following expression describes the quasilocal energy \eqref{QLEnergyIntegrated} in both cases
\begin{equation}\label{QLEnergySchw}
	E=\frac{r}{G}\,\left[1-\epsilon\sqrt{\epsilon\left(1-\varphi\right)\left(1-{\frac{r_s}{r}}\right)}\right]\ .\end{equation}

We have written the expressions above in such a way that taking $\varphi=0$ and $\epsilon=1$ for $r>r_s$ we get the quasilocal energy 
of the black hole as a whole as seen from outside\footnote{That is from the reference system of an observer at rest at $r = \infty$.}. The
quasilocal energy inside an $r=$ constant surface placed outside the event horizon is given by
\begin{equation}\label{QLESchwOut}
	E= \frac{r}{G}\,\left(1-\sqrt{1-\frac{r_s}{r}}\right)\ .\end{equation}
Assuming the existence of a BEC with $\varphi\neq0$ and that inside the 
horizon the $r$ coordinate becomes timelike with $\epsilon=-1$
\begin{equation}\label{QLESchwIn}
  E= \frac{r}{G}\,\left( 1+\sqrt{1-\varphi}\sqrt{\frac{r_s}{r}-1}\right)\  .\end{equation}
These expressions were derived in \cite{BrownYork} in the case $\varphi=0$.

\begin{figure}[!t]\begin{center}
	\includegraphics[scale=1.2]{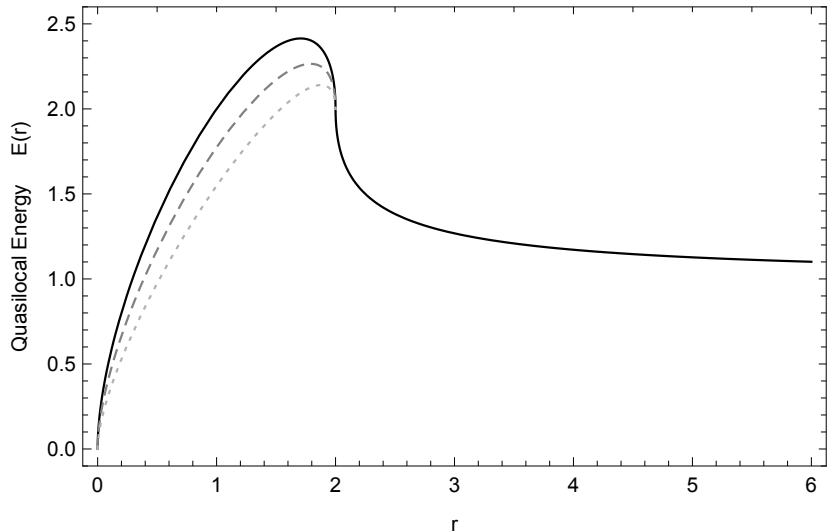}\vspace{40pt}
	\caption{Quasilocal energy for a Schwarzschild black hole computed inside as well as outside the event horizon. Both axes are in units of the mass $M$ and $G=1$. The event horizon is located at $2M$. In the inner region we compare a black hole in classical theory of general relativity where $\varphi=0$ (solid line) and various values for the condensate parameter, where we understand that a BEC structure is preferred minimizing the resulting energy of the object. We use as a comparison two values for the condensate: $\varphi=0.4$ dashed line and $\varphi=0.7$ dotted line.}\label{QLEPlotSchw}\end{center}\end{figure}

The quasilocal energy of the entire Schwarzschild metric is plotted in Figure \ref{QLEPlotSchw}.
This shows that a condensate structure lowers the energy with respect the familiar Schwarzschild metric.
The higher the value for $\varphi < 1 $ chosen, the lower the quasilocal energy will be.
Figure \ref{QLEPlotSchw} tells us that the total energy inside the event horizon (including gravitational energy as well as the 
energy of the $r = 0$ singularity) is twice the black hole mass
\begin{equation}
	E(r=r_s)=2M \end{equation}
and twice the energy of the total spacetime as can be seen from the asymptotic behaviour for the quasilocal energy at infinity.
        An expansion of \eqref{QLESchwOut} around infinity yields	
\begin{equation}
	\lim_{r\rightarrow\infty} E(r)\simeq \frac{r}{G}\,\left\{1-\left[1-\frac{r_s}{2\,r}+O\left(\frac{1}{r^2}\right)\right]\right\}=\frac{r_s}{2G}\equiv M\ ,\end{equation}
which is certainly an expected result.

The reinterpretation of a black hole as describing a graviton BEC does not modify three remarkable features of the quasilocal energy 
presented in \cite{Lundgren} that are perhaps not widely acknowledged. The first one is that the quasilocal energy is zero 
at the singularity. The infinity of the energy of the gravitational field 
at $r=0$ predicted by Newtonian gravity is removed. 
The second striking fact comes from the energy distribution inside the black hole. Inside the horizon,
the quasilocal energy has its maximum value at
\begin{equation}
	r_{max.}=GM\left(1+ \frac{1}{\sqrt{2-\varphi}}\right)\ .\end{equation}
In the limit of the classical theory, this maximum value is located at a radius equals to $\bigl(\,1+\frac{1}{\sqrt{2}}\,\bigr)GM$.
As the condensate parameter $\varphi$ increases, the maximum moves towards the event horizon and for the limiting value $\varphi=1$, it lies {\em exactly} at the horizon.
All this means that the black hole looks like an extended object where most of the energy seems to be `stored' not far from the horizon.
The third feature  is that the derivative of the quasilocal energy matches across the horizon, but is infinite there regardless of the value of $\varphi$.

Note that expanding the quasilocal energy and subtracting the BH mass M one gets for $r_s\ll r$
\begin{equation}
	E(r)-M = \frac{GM^2}{2r}+\dots.\end{equation}
The leading term is the classical gravitational self-energy of a shell of matter with total mass $M$ and radius $r$.

\subsection{Gravitational binding}\label{GravitationalBinding}
For $r\gg r_s$, if we denote by $E(r)\vert_M $ the quasilocal energy of a BH of mass $M$, 
\begin{equation}
	E(r)\vert_{M+m}-E(r)\vert_{M}\simeq m+G\,\frac{Mm}{r}+\frac32\, G^2\,\frac{M^2m}{r^2} +\dots=m\,\frac{\partial}{\partial_M} E(r) + {\cal O}(m^2)\ .\end{equation}
The first term corresponds to the gravitational potential energy of masses $m$ and $M$ separated by a distance $r$. 
It is also the energy (for large values of $r$) that it takes to separate the small mass $m$
from $M$ and take it to infinity.  
This leads us to interpret the quantity
\begin{equation}\label{potentialquasilocal}
	V(r)= 1-\frac{\partial}{\partial M}\left(\frac{r}{G}\sqrt{1-\frac{2GM}{r}}\right)=1-\frac{1}{\sqrt{1-\frac{2GM}{r}}}\end{equation}
as the binding potential energy associated to the quasilocal energy outside the BH horizon. Now, if we take
one step forward and extend these ideas to the BH interior, we will get 
\begin{equation}
V(r)= 1-\frac{\partial}{\partial M}\left(\frac{r}{G}\sqrt{1-\varphi}\sqrt{\frac{2GM}{r}-1}\right)
=1-\frac{\sqrt{1-\varphi}}{\sqrt{\frac{2GM}{r}-1}}\end{equation}
as the analogous potential energy inside the horizon.

\begin{figure}[!b]\begin{center}
	\includegraphics[scale=1.2]{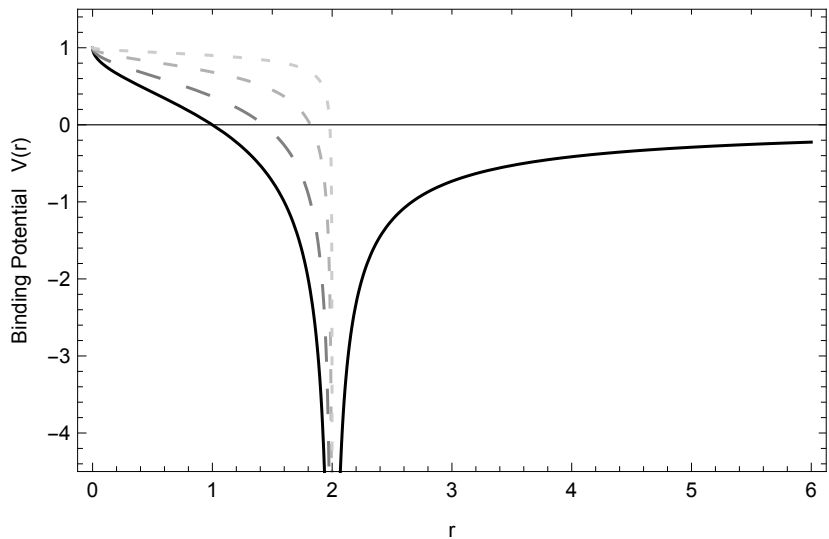}\vspace{40pt}
	\caption{Potential profile derived from Brown$-$York energy for Schwarzschild spacetime, with $M=1$ and various values for the condensate parameter $\varphi$. The solid line corresponds to $\varphi=0$. As $\varphi$ approaches the limiting value $\varphi=1$ the inner part of the potential tends to a square potential.}\label{Potential}\end{center}\end{figure}

We note that $V(r)$  is singular at $r=r_s$, but we
also note that this singularity is integrable. The following limits are readily found:
\begin{equation}
	\lim_{r\to 0}V(r)=1 \ ,\hspace{80pt}\lim_{r\to \infty}V(r)= 0\ .\end{equation}
A plot of the profile is presented in Figure \ref{Potential}. There we can see the potential that seems to
emerge from the Brown-York prescription for the quasilocal energy in the case where the background metric is Schwarzschild.
As we observe, the binding takes place in a region that is relatively close to the horizon, which is not at all 
unexpected after having seen the shape of the BH quasilocal energy profile.

There are several considerations that may follow from this potential profile if we accept the previous interpretation.
Notice that according to general relativity, a test particle of mass $m$
initially at rest at infinity would follow the trajectory described by the
geodesic (see e.g. \cite{rindler})
\begin{equation}
	\dot r^2= \left(1-\frac{r_s}{r}\right)^2\frac{r_s}{r}\ ,\end{equation}
where the dot means derivative with respect the coordinate time $t$ (i.e. not proper time). The radial momentum
will be  $p_r= m\gamma \dot r$. As the particle approaches $r=r_s$ from the outside, $p_r$ tends to zero as it is well
known. However, this does not take into account the emission of gravitational radiation
by the falling particle. If we take Figure \ref{Potential} at face value,  the particle
falls in the deep potential valley while emitting gravitational
radiation. However, this process should come to a stop because quantum 
mechanics should stabilize it at some point. Let us use a simple argument
to make some energy considerations (a similar argument works fine in the case of the hydrogen atom).

Indeed, assuming that $\Delta p_r \sim p_r$, according to the uncertainty principle 
$\Delta r > 1/p_r$. This leads to the following order of magnitude estimate:
\begin{equation}
	\Delta r \simeq \sqrt{\frac{r_s}{m}}\ .\end{equation}
This gives us the characteristic width of the potential and the approximate energy level.
Introducing the relevant units, for a BH of about 30 solar masses and considering the fall of an electron, this gives
$\Delta r \simeq 1 $mm. 
This result is surprising as it would imply that falling matter accumulates in a very thin shell on both sides of
 the horizon. The associated potential level would approximately be $\varepsilon/m \simeq 1-\sqrt{r_sm}$.

\begin{figure}[!b]\begin{center}
	\includegraphics[scale=1.2]{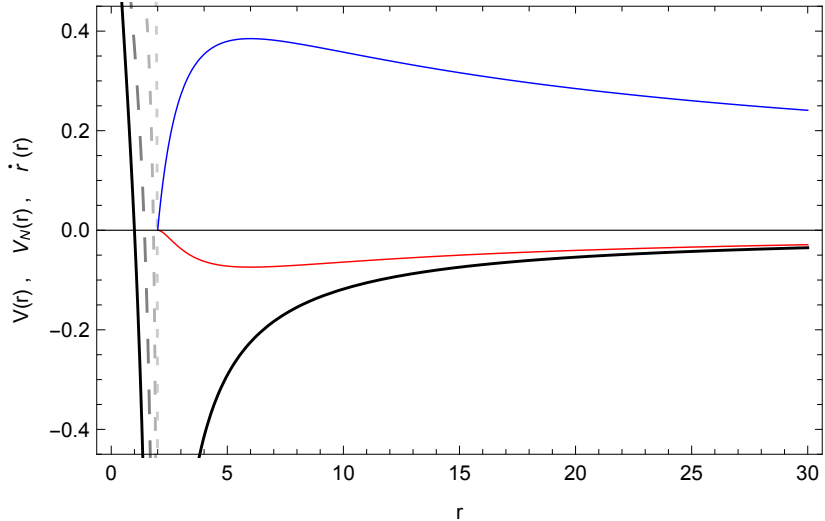}\vspace{40pt}
	\caption{Evolution of the velocity of a falling object as seen from a distant observer (blue curve). The maximum is $2c/(3\sqrt{3})$ and is attained at $r=3r_s$. We also plot the (fictitious) Newtonian potential that such an observer would `measure' from the fall of the object (red curve). In the same plot we show a zoom of the quasilocal potential profile derived from Brown$-$York energy for Schwarzschild spacetime, with $M=1$ presented in Figure \ref{Potential}. In the interior we plot the evolution of this magnitude as $\varphi$ gets close to its limiting value $\varphi=1$ and the potential approaches a step function.} \label{ParticleFalling}\end{center}\end{figure}

Figure \ref{ParticleFalling} may help us to understand the situation. In this figure we plot the velocity of a particle falling in a BH of mass $M=1$ following a radial geodesic, as seen from an observer at infinity. We also plot the Newtonian potential that such an observer would deduce from measuring the motion of the falling particle along its geodesic motion.
This (fictitious) potential agrees for the most part of the trajectory with the one derived from the quasilocal energy, which however is much deeper close to $r=r_s$. The difference between the two potentials should be observed by the distant reference point as gravitational wave emission. The total energy emitted should be approximately $m \sqrt{r_sm} \gg m$.
Obviously, only part of this energy is emitted outside the BH and eventually reaches a remote observer. It is to be expected that a good part of it remains inside the BH horizon.
This is a natural explanation for the graviton condensate, present when $\varphi\neq 0$, whose properties have been discussed in the previous sections.

Assuming for the sake of the discussion that (a) gravitons in the condensate are in a state close to the maximum packing condition,
i.e. have an energy $\sim 1/r_s$ and (b) that all the gravitational energy emitted by the falling particle is eventually stored in the BH interior, the capture by the BH of a particle of mass $m$ would create $n= (mr_s)^{\frac32}$ gravitons. This is a fairly large number so it is not unexpected that the continuous capture of matter by a BH brings the value of the graviton condensate close to 1, its limiting value,  $N\sim (M/M_P)^2$. At that point, most of the mass
of the BH is actually due to the graviton condensate: gravitational (quasi)local energy has turned into BH mass.
This simple argument gives strong plausibility to the thesis sustained by Dvali and G\'omez.

Certainly, along the process just described, $M$ and consequently the shape of the potential does not stay
 constant, so a detailed dynamical analysis
 would be required to understand this process more accurately.  Of course all the previous considerations hinge
 on the validity of the hypothesis behind the quasilocal energy principle.

It is actually interesting to see what happens when the value of the graviton BEC approaches the
 limiting value $\varphi=1$ beyond which the
 metric becomes intrinsically singular. The quasilocal potential becomes a constant (the quasilocal energy is a linearly rising function
 of $r$ in the BH interior in this limit). The result would apparently be a barrier right at the horizon location
 for falling particles. In this limit it would seem that gravitationally trapped objects would adhere to the BH external surface.

\section{BEC in Reissner–Nordström}\label{RRNN}
The Reissner–Nordström metric
\begin{equation}\label{MetricRN}
	\tilde{g}_{\mu\nu}\,dx^\mu dx^\nu=-\left(1-{\frac{r_s}{r}}+ \frac{r_Q^2}{r^2}\right)dt^2+\left(1-{\frac{r_s}{r}}+\frac{r_Q^2}{r^2}\right)^{-1}dr^2+r^2\,d\Omega^2\end{equation}
is a static solution to the 
Einstein–Maxwell field equations that is supposed to describe an spherically symmetric electrically charged black hole.
In order to introduce the electromagnetic interaction, the action \eqref{ActionFull} should be
modified by adding the following Lagrangian density
\begin{equation}
	\Delta S_{em}=\int d^4x\sqrt{-g}\,{\cal L}_{em}\ .\end{equation}
The corresponding electromagnetic stress–energy tensor is expressed in terms of the field strength $F_{\mu\nu}$ as usual
\begin{equation}\label{StressTensorElectro}
	T_{\mu\nu}=F_{\mu\sigma}F_\nu{}^\sigma-\frac{1}{4}F_{\sigma\tau}F^{\sigma\tau}\,g_{\mu\nu} \ .\end{equation}
The metric \eqref{MetricRN} describes the gravitational field of a charged, non-rotating, spherically symmetric body of 
mass $M$ and charge $Q$. In the above metric $r_Q^2=Q^2\,G$ is a characteristic length. 
The electromagnetic tensor \eqref{StressTensorElectro} describes the electromagnetic field in the outer region. 
The associated electric potential is $A_\mu=\left(Q/r,0,0,0\right)$.
 
This zone extends all the way from $r=\infty$ up to the (outer) event horizon
located at\footnote{As long as $2r_Q \leq r_s$; otherwise 
there can be no physical event horizon. Objects with a charge greater than their mass can exist in nature, but 
they cannot collapse down to a black hole unless they display a naked singularity.} 
\ben\label{EventHorizonRN}
	r_+=\frac12\left(r_s+\sqrt{r_s^2-4r_Q^2}\right),\een
where the metric becomes singular in these coordinates. There is another point where the metric presents a
singularity
\ben\label{EventHorizonRNCauchy}
	r_-=\frac12\left(r_s-\sqrt{r_s^2-4r_Q^2}\right).\een
This is called the inner horizon. In the region between horizons $r_- < r < r_+$ the
metric elements $\tilde g_{tt}$ and $\tilde g_{rr}$ exchange signs.

Let us entertain the possible existence of a condensate in this region. After adding the chemical potential term, 
the set of equations to solve are
\begin{equation}\label{EEqRN}
	G_{\mu\nu}(\tilde g+h)=\kappa T_{\mu\nu}+X_{\mu\nu}\ ,\end{equation}
where $T_{\mu\nu}$ is the electromagnetic tensor of the charged black hole \eqref{StressTensorElectro} and $X_{\mu\nu}$ 
is the chemical potential term defined as the LHS of equation \eqref{EEq}. We shall explore the existence of solutions of the type \eqref{MetricFullGeneric}. At the same time we redefine the electromagnetic tensor \eqref{StressTensorElectro} with a new unknown factor (this can be understood as a renormalization of the charge $Q$)
\be
	F_{\mu\nu}\quad\rightarrow\quad C\,F_{\mu\nu}\ .\ee

It can be seen that a family of solutions describing a condensate exists. Indeed from the angular equations we get
\begin{equation}\label{EinEqRNAngular}
	(1-\varphi)\,\frac{Q^2}{r^4}=-\, C^2\,g^{tt}g^{rr} \,\frac{Q^2}{r^4}\ ,\end{equation}
where the effective metric ($g=\tilde g +h$) is used to raise and lower indices. Because $\tilde{g}^{\,tt}\,\tilde{g}^{\,rr}=-1$
this equation defines the relation of the new  constant $C$  to the value of the condensate
\begin{equation}\label{C}
	C^2=\frac1{1-\varphi}\ .\end{equation}
On the other hand we have the sector where the chemical potential acts; the $(t,r)-\,$sector. The resulting equation is
\begin{equation}\label{RNEinEqTR}
	-\left(1-\varphi\right)\frac{Q^2}{r^4}-\frac{\varphi}{r^2}=\mu\,\left(1-\varphi\right)^2\varphi+C^2\,g^{tt}g^{rr}\,\frac{Q^2}{r^4}\ .\end{equation}
The value for $C$ in \eqref{C} makes the terms with $Q$ to cancel and this equation exactly reproduces the one
obtained in the Schwarzschild case in eq. \eqref{GPEquation}; hence the chemical potential yields
\be
	\mu r^2=-\frac{1}{(1-\varphi)^2}\ .\ee
As we have seen in previous cases, the value of the BEC is in principle arbitrary $0<\varphi<1$ and 
its relation to the chemical potential 
remains unchanged with respect to the Schwarzschild case.

All obvious limiting cases behave as expected. If we turn off the charge, $Q=0$, the field equations and their 
corresponding solutions reduces to the uncharged Schwarzschild case. Also, if we make the chemical potential 
to vanish, 
the condensate disappear and we are left with the classical Reissner–Nordström vacuum solution.  
Maxwell equations in vacuum are still satisfied in this solution for any arbitrary $Q$
\be\begin{split}
	&{\cal D}^\nu F_{\mu\nu}=0\ ,\\
	&F_{\nu\rho,\mu}+F_{\rho\mu,\nu}+F_{\mu\nu,\rho}=0\ .\end{split}\ee
The electromagnetic energy $(E^2+B^2)/2$ is scaled by the factor $C^2= \frac{1}{1-\varphi}$ if $\varphi\neq 0$.
This is indeed equivalent
to scaling the charge. Notice that the redefinition of the charge does not change the background metric, nor, consequently,
the location of the two horizons.

Reissner-Nordstr\"om is not a particularly physically relevant solution as the existence of a macroscopic electrically
charged BH is very unlikely, so its interest is mostly theoretical as it would imply an accumulation of charged matter.
Taking this into account one can possibly understand the modification of the charge implied by the redefinition $Q\to CQ$ as
the consequence of the fact that the condensate density increases too.
This redefinition makes $Q$ to diverge when $\varphi \to 1$.

\subsection{Quasilocal energy in Reissner–Nordström}
We shall use again the results in Section \ref{QLE}. In this case, the radial metric element is 
\begin{equation}\label{fRN}
	f(r)=\sqrt{1-\varphi}\sqrt{\epsilon\left(1-\frac{r_s}{r}+\frac{r_Q{}^2}{r^2}\right)}\ .\end{equation}
In the outer horizon the same change of sign of $\epsilon$ as in the Schwarzschild case takes place. 
As we move inwards and cross the inner horizon the signature of the $t$ and $r$ coordinates are exchanged again, 
and $t$ recovers its timelike character.

Including the electromagnetic field into the theory will not change the definition of the quasilocal energy. 
The quasilocal energy is meant to measure the gravitational energy associated to a specific geometry, and so 
only the gravitational action is important. Of course, the addition of a new field changes the metric
and this is the way that adding the electromagnetism influences the quasilocal energy.

The quasilocal energy in this case is given by 
	\be
	E(r)=\frac{r}{G}\left[1-\epsilon\sqrt{1-\varphi}\sqrt{\epsilon\left(1-\frac{2m}{r}+\frac{Q^2}{r^2}\right)}\right]\ ,\ee
with $\epsilon=1$ when $t$ is timelike and  $\epsilon=-1$ when the temporal coordinate is spacelike.
A striking feature is that the energy becomes negative for $r< Q^2/2M$. This radius is always inside the inner horizon. 
The value of the quasilocal energy at the singularity is $E(r=0)=-\abs{Q}$. 
The singularity at $r=0$ has the electric field of a point charge, 
and so, using just classical electromagnetism, its self energy should diverge.

\begin{figure}[t]\begin{center}
	\includegraphics[scale=1.2]{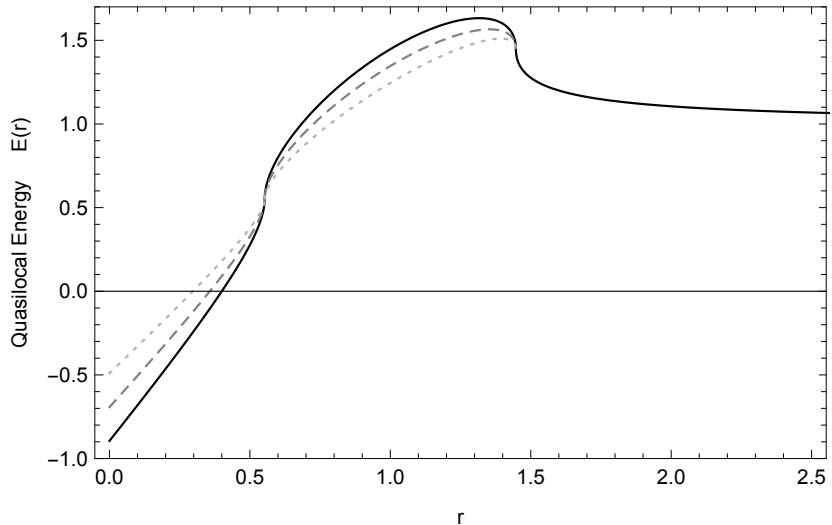}\vspace{40pt}
	\caption{Quasilocal energy of a Reissner–Nordström black hole, with a charge given by $Q^2=0.8M^2$. Both horizons are clearly identified by the divergence of the derivative of the energy. Inside the external event horizon we compare the classical theory where the wave function is null, $\varphi=0$, with a condensate structure with the same values for $\varphi$ used in the Schwarzschild case. As we can see from the plot, inside the Cauchy horizon (where the $t$ and $r$ coordinates exchange signs again), the condensate structure is not favoured any more; hence the theory dictates a ring condensate between both horizons. Both axes are in units of the mass $M$ of the black hole.}\label{QLE.RN}\end{center}\end{figure}
	
In Figure \ref{QLE.RN}  the quasilocal energy for these type of black holes is presented. As we see, in the regions 
$r> r_+$ and $r<r_-$ the usual (no condensate) solution is energetically preferred, while in the
region between horizons the opposite behaviour is taken place and a condensate is viable. In particular, inside the inner horizon
there should not be a condensate.
We conclude that for this case the condensate has the structure of a spherical shell. According to the proponents of the
quasilocal energy, this energy includes the contribution of all fields.

The conclusions that can be drawn are in line with those discussed in the previous section for a Schwarzschild BH.

\newpage
\section{BEC in de Sitter}\label{dS}
It is quite peculiar for reasons that we will discuss below that also in this case the presence of an event horizon seems associated to a possible BEC. Let us repeat the previous calculations. In a de Sitter universe a cosmological term must be included to the Hilbert-Einstein part of the action \eqref{ActionFull}
\begin{equation}
	S_{dS}=M_P^2 \int d^4x\sqrt{-g}\,(R-2\Lambda)\ . \end{equation}
The Einstein field equations with cosmological constant and chemical potential in this case are 
\begin{equation}
	G_{\mu\nu}=-\Lambda g_{\mu\nu} + X_{\mu\nu}\ ,\end{equation}
giving rise to the following maximally symmetric solution
\begin{equation}\label{MetricdS}
	ds^2=-\left(1-\frac{\Lambda}{3}\,r^2\right)\,dt^2+\left(1-\frac{\Lambda}{3}\,r^2\right)^{-1}\,dr^2+r^2\,d\Omega^2\end{equation}
that possesses a cosmological horizon at 
\begin{equation}
	r_\Lambda=\sqrt{\frac3\Lambda}\ .\end{equation}
Searching for solutions with a non-zero chemical potential, we redefine the cosmological constant term
with the help of a new constant $C$:   $\Lambda \rightarrow C\,\Lambda$, while the metric
is modified to
\begin{equation}\label{MetricFullGenericDS}
  g_{\mu\nu}=\text{diag}\left(\frac{1}{1-\varphi}\,\tilde g_{tt}\,;\,\frac{1}{1-\varphi}\,\tilde g_{rr}\,;\,\tilde g_{\theta\theta}\,;\,\tilde g_{\phi\phi}\right), \end{equation}
as in the Schwarzschild and Reissner-Nordstr\"om cases.

Any of the two angular components of the Einstein's equations fixes this constant with respect to the condensate as
\begin{equation}
	C=1-\varphi \ .\end{equation}
This election for $C$ makes the remaining field equations to take the same form as in \eqref{GPEquation}; 
the cosmological contribution inside the Einstein tensor cancel out with the modified cosmological 
term. As the equation is the same, the structure for the chemical potential 
remain unchanged with respect to the Schwarzschild case \eqref{ChemPot}. That is
\be
	\mu(r) = \frac{\mu_0}{r^2}\ .\ee
In the visible part of our universe there is no condensate. This case is fairly obvious as at $r=0$ the metric is Minkowskian. At $r=r_\Lambda$ there is a discontinuity in the chemical potential. There is a peculiar
situation here: contrary to the Reissner-Nordstr\"om case, where the electromagnetic field is enhanced by the condensate structure,
the cosmological constant term is screened by the condensate structure. The presence of the graviton condensate decreases
the value of the cosmological constant which becomes zero for the limiting value $\varphi=1$.

\subsection{Quasilocal energy in de Sitter}

\begin{figure}[!b]\begin{center}
	\includegraphics[scale=1.2]{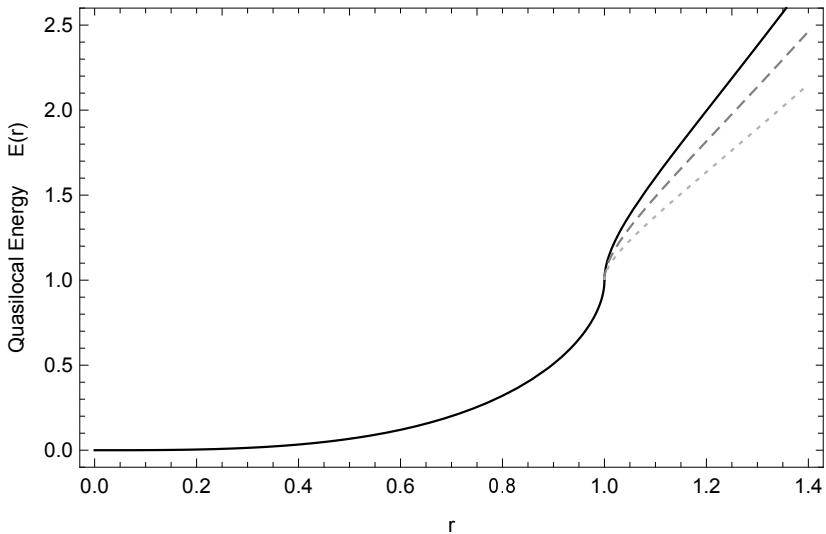}\vspace{40pt} \caption{Quasilocal energy for de Sitter spacetime. The axes are in units of the cosmological horizon, i.e. $r_\Lambda=1$. Beyond the horizon we present the curves for the classical solution in solid line and two different condensate values, $\varphi=0.2$ and $\varphi=0.4$, with dashed line and dotted respectively.} \label{QLEdeSitter}\end{center}\end{figure}
	
The fundamentals of the previous  derivations of the quasilocal energy do not change if a cosmological constant is considered. Now $f(r)$ in \eqref{Metric} is given by
	\ben
	f(r)=\sqrt{1-\varphi}\sqrt{\epsilon\left(1-\frac{\Lambda}{3}\,r^2\right)}\ ,\een
where $\varphi$ is zero in our visible universe. When $\epsilon=1$, $g_{tt}$ is negative and $t$ 
is a timelike coordinate. At a horizon, $g_{tt}$ and $g_{rr}$ exchange signs and so $\epsilon=-1$. Then
	\ben\label{QLEdS}
	E(r)=\frac{r}{G}\left[1-\epsilon\sqrt{1-\varphi}\sqrt{\epsilon\left(1-\frac{\Lambda}{3}\,r^2\right)}\right].\een
The quasilocal energy for a de Sitter metric is shown in Figure \ref{QLEdeSitter}. 
As expected, the energy continually grows with increasing $r$. The horizon forms when the energy 
at the surface is larger than $E(r_c)=r_\Lambda/G$. 
As in the Schwarzschild case, the formation of a condensate diminishes the energy with respect to the vacuum case, so it 
seems that the condensate structure is favorable against the vacuum itself.

\begin{figure}[!h]\begin{center}
	\includegraphics[scale=1.2]{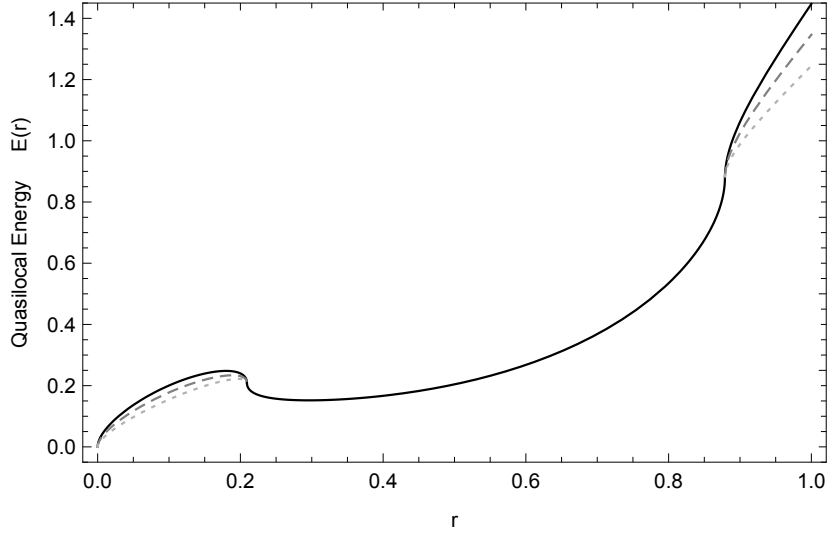}\vspace{40pt} \caption{Quasilocal energy for a Schwarzschild black hole embedded in a de Sitter space. The units are such that $r_\Lambda=1$. The black hole mass is unrealistically large, $r_s=0.1\,r_c$ to make details notorious. Beyond both horizons we plot a condensate structure for comparison, with $\varphi=0.2$ and $\varphi=0.4$ in dashed and dotted lines respectively.} \label{QLESchw-deSitter}\end{center}\end{figure}

Just for completeness the solution combining the Schwarzschild and de Sitter spacetimes can be easily computed too. 
The quasilocal energy is plotted in Figure
\ref{QLESchw-deSitter}.
Both horizons, the black hole horizon and the cosmological horizon, can be directly seen in the curve. 
In both points even though the derivative of the energy matches across the horizon, it is divergent.
Also in both points, the horizon forms when the quasilocal energy crossed the value $r/G$.

The above picture provides a criterion to discern whether a condensate can be formed or not. 
Any time the radial coordinate transform its character from spacelike to timelike, when crossing a horizon, 
a condensate structure is energetically favoured with respect to the vacuum solution.

\vspace{30pt}

\subsection{The case of the de Sitter horizon}
Before entering in a more detailed discussion about possible physical interpretations in the case
of the cosmological horizon in de Sitter, let us 
analyze this spacetime but using an inverse radial coordinate
\begin{equation}
	\left\{\begin{split}
	&\quad r=\frac{1}{z} \\ 
	&\quad dr=-\frac{1}{z^2}\,dz\ .
	\end{split}\right.\end{equation}
With this change, the metric reads as
	\begin{equation}\label{MetricdSz}\begin{split}
	ds^2=-\left(1-\frac{\Lambda}{3}\,\frac{1}{z^2}\right)\,dt^2+\left(1-\frac{\Lambda}{3}\,\frac{1}{z^2}\right)^{-1}\,\frac{1}{z^4}\,dz^2+\frac{1}{z^2}\,d\Omega^2\end{split}.\end{equation}
This change of variables maps the cosmological horizon to a concentric sphere centered at $z=0$. The Einstein tensor is constant in $r$ coordinates, then it is not changed by this change of variables and the equations of motion remains the same as well.
Let us now consider the geodesics in this spacetime for straight trajectories characterized by constant angles
$\theta$ and $\phi$. Some Christoffel symbols used to compute geodesics are
\begin{equation}\begin{split}
	&\Gamma_{tt}^z=\frac{3\,\Lambda\,z^2-\Lambda^2}{9\,z}\\
	&\Gamma_{tz}^t=\frac{\Lambda}{3\,z^3-\Lambda\,z}\\
	&\Gamma_{zz}^z=-\frac{6\,z^2-\Lambda}{3\,z^3-\Lambda\,z}\ .\ \end{split}\end{equation}
The geodesic equation is 
\begin{equation}
\frac{d^2z}{dt^2}=-\frac{3z^2-\Lambda}{9z}\,\Lambda+\frac{6z^2+\Lambda}{3z^3-\Lambda z}\left(\frac{dz}{dt}\right)^2\ .
\end{equation}
This equation is virtually identical in structure to the one corresponding to a particle moving towards 
the horizon in a Schwarzschild metric with vanishing angular momentum and, exactly like in that case, the
observer sees the particle to slow down as it approaches the horizon that ---at least classically--- never 
crosses (according to the observer at infinity in the case of Schwarzschild, at $r=0$  in the case of 
de Sitter, i.e. at $z=\infty$).

Continuing with this analogy we derive the quasilocal energy using the $z$ coordinate where
the space beyond the de Sitter horizon is mapped onto a sphere of radius $z_\Lambda= 1/r_\Lambda$.
It is given by
\begin{equation}\label{quasilocaldesitterz}
	E(z)=\frac{1}{z}\,\left(1-\sqrt{1-\frac{\Lambda}{3}\frac{1}{z^2}}\right)\end{equation}
for $z> z_\Lambda$ and
\begin{equation}
	E(z)=\frac{1}{z}\,\left(1+\sqrt{1-\varphi}\,\sqrt{\frac{\Lambda}{3}\frac{1}{z^2}-1}\right)\ \end{equation}
for $z< z_\Lambda$.
The derivation for this expression was made directly using the inverse $z$ coordinate.
If we go back to the radial coordinate  $z\rightarrow r$ we obtain the expression \eqref{QLEdS} but with a difference of a
global minus sign.

Radial geodesics (in the usual coordinate $r$) can be obtained easily from the expression
\begin{equation}
  \dot r^2= \left(1-\frac{\Lambda r^2}{3}\right)^2\frac{\Lambda r^2}{3}\ ,
\end{equation}
where $\Lambda r^2/6$ would be the analogous of the Newtonian potential. Note that the usual
interpretation that one element of vacuum repeals each other, making the expansion of the universe obvious, is very intuitive in
Friedmann-Robertson-Walker coordinates, but is not so obvious in the static coordinates used in this section.

In spite of the formal similarity between the expressions for the quasilocal energy (\ref{quasilocaldesitterz}) and
(\ref{QLESchwOut}), the situation is different because it does not seem possible to derive a potential such as the one
in (\ref{potentialquasilocal}). The reason is that (\ref{quasilocaldesitterz}) is mass-independent. 

As a consequence, the argument that was presented in the case of a Schwarzschild black hole, where it was argued that 
the presence of the potential well implied a substantial amount of gravitational radiation that could be stored in 
the inner part of the black hole in form of a graviton condensate does not hold. In the present case the potential well appears 
to be absent and consequently there is no energy loss in form of gravitational radiation.

Intuitively, the de Sitter horizon appears to be somewhat different from the one associated to a Schwarzschild BH (or to a
Reissner-Nordstr\"om for that matter). To the external observer at infinity, the horizon at $r=r_s$ has a
very clear meaning and objective existence. On the contrary, in the de Sitter spacetime the horizon is dependent
on the observer situation (i.e. where the $r=0$ point is assumed to be). However as discussed in
some detail in the seminal paper of Gibbons and Hawking \cite{GibbonsHawking} this might not pose any conceptual problem
in principle. It is only  measurable effects by an observer what matters. 

As it is well known, there is a temperature associated to the de Sitter horizon, given by \cite{GibbonsHawking}
\begin{equation}
	T=\frac{1}{2\pi}\sqrt{\frac{\Lambda}{3}}\ .\end{equation}
This raises a puzzling possibility: matter falling into the de Sitter horizon in the inverse coordinate $z$; actually accumulating
close to the horizon in the static coordinates $(r,t)$. In the Schwarzschild case we concluded that this would most likely imply
a growth in the value of the `graviton' condensate $\varphi$. 
If this were the case, we immediately notice that because of the change in
the cosmological constant beyond the horizon $\Lambda \to C\Lambda$ with $C= 1-\varphi$ the vacuum energy would decrease with time.
However, this does not affect neither the position of the horizon, as the change in the metric is merely a multiplicative  
factor in the temporal and radial components, nor the value of the cosmological constant in the visible part of the 
universe.

However, the fact that there is no gravitational binding potential of the form (\ref{potentialquasilocal}) obtained in the
Schwarzschild case, would imply that there is no way to create a gravitational BEC and therefore, in spite of being
energetically favourable, the value of the condensate should be zero and consequently there would be no change in the cosmological 
constant neither inside nor outside the de Sitter horizon.

\section{Conclusions and outlook}

In the present work new insights on the possibility of identifying black holes with Bose-Einstein condensates of gravitons are proposed.
In a previous work, we derived a consistent picture in the context of Schwarzschild black holes in \cite{AEG} and now we extend the analysis to
charged black holes (Reissner-Nordstr\"om) and the de Sitter cosmological horizon. We have found that in all these situations a graviton condensate is possible in
the classically inaccessible zone when
a chemical potential term is appropriately introduced in Einstein's equations. This seems to suggest that the existence of an horizon
is intimately linked to the existence of such condensates.

In order to discern whether this possibility is actually realized in nature, we analyze how the notion of quasilocal energy 
in the context of the classical theory of general relativity can be extended so as to encompass the condensate description.
Even though there is no total consensus on the definition of a quasilocal energy, we consider the Brown-York prescription\cite{BrownYorkMath}
as being adequate for our purposes and physically well-motivated in the present context.
If this definition for the energy it chosen, it is found ---and this is one of our main results--- that a graviton BEC is energetically favourable
for all the studied cases in the regions beyond the horizon.

The possibility of defining a quasilocal energy entails the possibility of deriving a quasilocal potential describing the binding of falling matter to
a black hole. Its well-like structure, makes the binding to take place around a fairly thin shell located at the event horizon.
This seems to indicate that matter falling into the black hole accumulates at both sides of the horizon, but always in its vicinity.
This confers to the black hole a fuzzy boundary. When matter falls and get trapped by this potential, gravitational wave emission is necessarily
produced on energy conservation grounds. Part of this energy is emitted towards infinity, however it is expected that a large fraction is trapped
in the inner region. This seems to be a very plausible explanation on how the condensate of gravitons forms and the value of the graviton
condensate order parameter $\varphi$ increases. As explained in section \ref{GravitationalBinding}, order of magnitude considerations indicate
that practically most of the black hole mass may be stored in the form of a  graviton BEC. Besides, the unceasing capture of
gravitons may indicate that the condensate value should be close to its limit value, $\varphi\rightarrow1$. This statements gives strong
plausibility to the original proposal of Dvali and G\'omez\cite{DG1}.

Extension of these considerations to the Reissner-Nordstr\"om case is straightforward.
However, is not always  possible to define a significant quasilocal potential for every spacetime. For instance, in de Sitter we have been unable to
do so, this concluding that while it is energetically viable, a graviton BEC is actually not present.

The validity ---quite plausible in our opinion--- of the previous picture brings about many interesting points. One of them is
that the horizon, while being from the metric point of view, quite well defined, becomes necessarily fuzzy due to quantum effects.
A second consequence is that matter falling in the black hole does not get `crunched' by the singularity at $r=0$. When it comes to the quasilocal
energy, it is perfectly regular at the origin, and so is the associated potential (it tends to a constant).
In the classical theory,
the location of the horizon is defined as a limit for timelike world lines to exist. Points where 4-velocity turns null form a `one way' spatial surface.
Here, following our interpretation, we see that matter is captured by the quasilocal potential and certainly cannot escape, except for the occasional
thermal fluctuation. However, no loss of information would occurs here, except for the usual thermodynamic irreversibility. If the mass of the black hole decreases
for some reason and $r_s\to 0$, the potential becomes progressively shallower and matter stored in its potential well can escape. Needless to say that one
mechanism whereby the mass of the black hole could decrease is through evaporation. This is an intrinsically quantum mechanism that is
accurately described in the usual way.
However, if a graviton condensate is present, and we have argued that this is very likely with a value close to the limiting value, one
should take into account that Bose-Einstein condensates are not localized objects and therefore some amount of leaking should be present. In their original
proposal Dvali and G\'omez sustained the point of view that Hawking radiation could be understood in this way. From the point of view that most
of the black hole mass is stored in the form of the condensate, this is a likely possibility, but we have not considered this issue in the present work (see
however \cite{AEG}).

In spite of the previous considerations, the paradox of loss of information continues to be present. In the present description, which is a continuation of
\cite{AEG}, one has to assume the pre-existence of a black hole, however small, bringing again the familiar issues about trapped surfaces, etc. However, the
problem seems somewhat ameliorated due to the mechanisms described here.

\section*{Acknowledgments}
We acknowledge the financial support of the research grants FPA2016-76005-C2-1-P and MDM-2014-0369. 
The work of J.A. is partially supported by grants
Fondecyt 1150390 and CONICYT-PIA-ACT14177 (Government of Chile). L.G. is supported by an FPI from MINECO (Spain), grant BES-2014-067939.

\begin{appendices}
\section{\hspace{-16pt}. Quasilocal energy }
\label{quasilocalenergy}
We will deal with a compact spacetime domain $M$ described by the metric $g_{\mu\nu}$ that in our static and spherically symmetric cases, gets
the structure written in \eqref{Metric}. Time is globally defined and provides a foliation $\Sigma$. This family of space-like hypersurfaces
has a future-directed timelike unit normal $u_\mu=-N\delta_\mu^t$.
Space coordinates $i=(r,\theta,\phi)$ adapted to this foliation can be introduced, leading to
the metric tensor $\gamma_{ij}$, given by \eqref{MetricSigmaSphSym}, and the extrinsic curvature $\Theta_{ij}$ on $\Sigma$.

Each hypersurface $\Sigma$ has a spatially closed two-boundary $\partial\Sigma=B$ defined by its normal vector \eqref{Normal} in $\Sigma$.
The induced two-metric is written as $\sigma_{ij}=\gamma_{ij}-n_in_j$ and has the geometry of a sphere.  
The corresponding extrinsic curvature $K_{\mu\nu}$ has already been defined in \eqref{ExtrinsicCurvatureB}.
The time history of these two-surface boundaries $B$ is the timelike three-surface boundary $^3B=B\times[t',t'']$ of $M$, defined by a spacelike normal
$n^\mu$ pointing outwards. Intrinsic coordinates $i=(t,\theta,\phi)$ can also be introduced with the associated metric  $\overline\gamma_{ij}$. 
In what follows, the overlined notation refers to tensors defined on $^3B$. For instance, the
extrinsic curvature associated with the three-boundary as embedded in $M$ is $\overline\Theta_{ij}$.

In order to apply the action principle to the domain $M$ we need to include all boundary terms: 
	\begin{equation}
	  S=\frac1{2\kappa}\int_Md^4x\sqrt{-g}R+\frac1{\kappa}\int_{\Sigma}d^3x\sqrt{\gamma}\ \Theta\,\bigg\vert_{t'}^{t''} -\frac1{\kappa}\int_{^3B}d^3x\sqrt{-\overline\gamma}\ \overline\Theta+\frac1\kappa\int_{^3B}d^3x\sqrt{-\overline\gamma}\ \overline{\overline\Theta}\ .\end{equation}
The last term is the Gibbons-Hawking normalizing factor \cite{Gibbons}, where $\overline{\overline\Theta}$ corresponds to the trace of the extrinsic curvature
of the three-boundary, but as embedded in a flat background four-geometry (the spacetime in which we want to obtain zero quasilocal quantities).
This term is usually interpreted as our freedom to shift the zero point of the energy. 

Varying the action we get
\begin{equation}\label{ActionVariationGR}\begin{split}
	\delta S=&\ \Big(\begin{gathered}\text{ equations of motion }\\[-1ex] \text{terms}\end{gathered}\Big)+ \int_\Sigma d^3x\ P^{ij} \delta \gamma_{ij}\Bigr\vert_{t_i}^{t_f}+ \int_{^3B} d^3x\ \left(\pi-\pi_0\right)^{ij}\delta\overline\gamma_{ij} \ , \end{split}\end{equation}
        where $P^{ij}$ and $\pi^{ij}$ are the momenta canonically conjugate to the corresponding metrics of each of the two submanifolds (the $\pi_0^{ij}$ term is conjugate to the flat metric and is simply a shift due to the Gibbon-Hawking normalizing factor; fixing this term means to choose a ‘reference configuration’). The guiding principle for obtaining the quasilocal energy is the analogy between the latter variation and the action variation in nonrelativistic mechanics:
\begin{equation}\label{ActionVariationNR}\begin{split}
	\delta S=&\ \Big(\begin{gathered}\text{ equations of motion }\\[-1ex] \text{terms}\end{gathered}\Big)+p\,\delta q -H\,\delta t\ . \end{split}\end{equation}
When restricted to classical solutions, the equations of motion vanish. We have the correspondence
\begin{equation}\label{HJEq}\left\{\begin{split}
	&p_i=\frac{\partial S}{\partial q_i} \\
	&H=-\frac{\partial S}{\partial t} \end{split}\right.\qquad\Leftrightarrow\qquad \left\{\begin{split}
	&P^{ij}=\frac{\delta S}{\delta \gamma_{ij}} \\
	&(\pi-\pi_0)^{ij}=\frac{\delta S}{\delta\overline\gamma_{ij}}\end{split}\right. \ .\end{equation}
The generalization is quite direct. The three-metric $\overline\gamma_{ij}$ provides the metrical distance between all spacetime intervals in the boundary manifold $^3B$ (including time between spacelike surfaces); therefore the notion of energy (the one equation for the Hamiltonian) in nonrelativistic
        mechanics is generalized to a stress energy momentum defined on $^3B$ that characterizes the entire system (gravitational field, chemical potential fields and any matter fields and/or cosmological constant). In accordance with the standard definition for the matter stress tensor $T^{\mu\nu}$, the following surface stress tensor is defined
	\begin{equation}\label{StressTensorSurface3B}
	\tau^{ij}\equiv\frac2{\sqrt{-\overline\gamma\,}}\frac{\delta S}{\delta\overline\gamma_{ij}}=\frac2{\sqrt{-\overline{\gamma}\,}}\left(\pi^{ij}-\pi^{ij}_0\right)\ .\end{equation}

An important feature of this stress tensor is apparent when considering two concentric spherical surfaces $B_1$ and $B_2$. In the limit where $B_1$
        and $B_2$ approach each other, the total surface stress-energy-momentum result in $\tau^{ij}=2/\sqrt{-\overline\gamma}(\pi_2^{ij}-\pi_1^{ij})$
        since the reference terms $\pi_0^{ij}$ cancel between each other. This tensor embodies the well-known result of Lanczos and Israel
        in general relativity \cite{Lanczos-Israel}, which relates the jump in the momentum $\pi^{ij}$ to the matter stress tensor of the surface layer.
        In the infinitesimally thin layer limit, the geometries of each side of the layer coincide and there is no gravitational contribution to $\tau^{ij}$.
        The direct physical implication of this result is the widely known absence of a local gravitational energy momentum \cite{MisnerThorneWheelerLibro}.

Therefore, from the previous discussion
\begin{equation}\label{MomentumTensors}
	P^{ij}=\frac1{2\kappa}\sqrt{\gamma}\left(\Theta\,\gamma^{ij}-\Theta^{ij}\right)\ ,\hspace{40pt}\pi^{ij}=-\frac1{2\kappa} \sqrt{-\overline\gamma}\left(\overline\Theta\,\overline\gamma^{\,ij}-\overline\Theta^{\,ij}\right)\ ,\end{equation}
are canonically conjugate to the metrics $\gamma_{ij}$ and $\overline\gamma_{ij}$ respectively.
The normal and tangential projections of $\tau^{ij}$ on the two-surface $B$ defines the proper energy $\varepsilon=u_iu_j\tau^{ij}$, momentum $j_a=-\sigma_{ai}u_j\tau^{ij}$, and spacial-stress $s^{ab}=\sigma_i^a\sigma_j^b\tau^{ij}$ surface densities. We are interested in the first magnitude (further details can be found in \cite{BrownYork}); integrating the energy density along $B$ leads us directly to the definition of the quasilocal energy given in \eqref{QLEnergy}
	\begin{equation}
	E=\int_Bd^2x\ \sqrt{\sigma}\,\varepsilon=\frac1\kappa \int_Bd^2x\ \sqrt{\sigma}\,\left(K-K_0\right)\ . \end{equation}
The energy result in the subtraction of the total mean curvature of $B$ as embedded in $\Sigma$ with the total mean curvature of $B$ as embedded in a Minkowskian reference frame, times the inverse of the Einstein's gravitational constant.

\end{appendices}


\begin{thebibliography}{99}

\bibitem{DG1} 
G. Dvali, C. Gomez and S. Zell, {\it Quantum Break-Time of de Sitter}, J. Cosmol. Astropart. Phys. {\bf 06}, 028 (2017); arXiv:1701.08776.

G. Dvali and C. Gomez, \textit{Black Holes as Critical Point of Quantum Phase Transition}, Eur. Phys. J. C \textbf{74}, 2752 (2014); arXiv:1207.4059.

G. Dvali and C. Gomez, \textit{Black Hole’s 1/N Hair}, Phys. Lett. \textbf{719}, 419 (2013); arXiv:1203.6575.

G. Dvali, D. Flassig, C. Gomez, A. Pritzel and N. Wintergerst, \textit{Scrambling in the Black Hole Portrait}, Phys. Rev. D \textbf{88}, 124041 (2013); arXiv:1307.3458.

G. Dvali and C. Gomez, \textit{Black Hole’s Quantum N--Portrait}, Fortsch. Phys. \textbf{61}, 742 (2013); arXiv:1112.3359.

G. Dvali and C. Gomez, \textit{ Landau-Ginzburg Limit of Black Hole’s Quantum Portrait: Self Similarity and Critical Exponent}, Phys. Lett. B \textbf{716}, 240 (2012); arXiv:1203.3372.



\bibitem{Radiation}
M.K. Parikh and Frank Wilczek, {\it Hawking Radiation as Tunneling}, Phys. Rev. Lett. {\bf 85}, 5042 (2000); hep-th/9907001.

S.W. Hawking, \textit{Particle Creation by Black Holes}, Commun. Math. Phys. \textbf{43}, 199 (1975).



\bibitem{Entropy} 
J.D. Bekenstein, \textit{Black Holes and Entropy}, Phys. Rev. D \textbf{7}, 2333 (1973).

J.D. Bekenstein, \textit{Generalized second law of thermodynamics in black-hole physics}, Phys. Rev. D \textbf{9}, 3292 (1974).

S.W. Hawking, \textit{Black holes and thermodynamics}, Phys. Rev. D \textbf{13}, 191 (1976).



\bibitem{NoHair} 
J.D. Bekenstein, Phys. Rev. Lett. \textbf{28}, 452 (1972).

J.D. Bekenstein, Phys. Rev. D \textbf{5}, 1239 (1972).

J.D. Bekenstein, Phys. Rev. D \textbf{5}, 2403 (1972).

C. Teitelboim, Phys. Rev. D \textbf{5} (1972) 294.

J. Hartle, Phys. Rev. D \textbf{3}, 2938 (1971).



\bibitem{InformationLoss}
S.W. Hawking, {\it Information Loss in Black Holes}, Phys. Rev. D \textbf{72}, 084013 (2005); hep-th/0507171.

S.W. Hawking, \textit{The Unpredictability of Quantum Gravity}, Commun. Math. Phys. \textbf{87}, 395 (1982).

For a review on information loss paradox, see, J. Preskill, \textit{Do Black Holes Destroy Information?}, International Symposium on Black Holes, Membranes, Wormholes, and Superstrings, Houston Advanced Research Center (1992); arXiv:hep-th/9209058.



\bibitem{AEG} 
J. Alfaro, D. Espriu and L. Gabbanelli, {\it Bose$-$Einstein graviton condensate in a Schwarzschild black hole}, Class. Quantum Grav. {\bf 35}, 015001 (2018); arXiv:1609.01639.



\bibitem{Germani}
F. Cunilleraa and C. Germani, {\it The Gross–Pitaevskii equations of a static and spherically symmetric condensate of gravitons}, Class. Quant. Grav. \textbf{35}, 105006 (2018); arXiv:1711.01282.

\bibitem{GPEquation} 
F. Dalfovo, S. Giorgini, L.P. Pitaevskii and S. Stringari, Theory of Bose-Einstein condensation in trapped gases. Rev. Mod. Phys. \textbf{71}, 463 (1999).

\bibitem{AEP} 
J. Alfaro, D. Espriu, D. Puigdom\`enech, \textit{Spontaneous generation of geometry in four dimensions}, Phys. Rev. D \textbf{86} 025015 (2012); arXiv:1201.4697.

J. Alfaro, D. Espriu, D. Puigdom\`enech, {\it The emergence of geometry: a two-dimensional toy model}, Phys. Rev. D \textbf{82}, 045018 (2010); arXiv:1004.3664.

\bibitem{Weinberg}
S. Weinberg, \textit{Cosmology} (Oxford Univ. Press, 2008).

\bibitem{Szabados}
L.B. Szabados, {\it Quasi-Local Energy-Momentum and Angular Momentum in General Relativity}, Living Rev. Relativ. {\bf 12}, 4 (2009).

\bibitem{BrownYorkMath}
J.D. Brown and J.W. York, {\it Quasilocal energy in general relativity}, in Mathematical Aspects of Classical Field Theory, edited by M.J. Gotay, J.E. Marsden, and V.E. Moncrief (American Mathematical Society, Providence, RI, USA, 1992), pp. 129–142.

\bibitem{YorkKAction}
J.W. York, {\it Boundary terms in the action principles of general relativity}, Found. Phys. \textbf{16}, 249 (1986).

J.W. York, {\it Role of Conformal Three-Geometry in the Dynamics of Gravitation}, Phys. Rev. Lett. {\bf 28}, 1082 (1972).

\bibitem{Braden}
H.W. Braden, J.D. Brown, B.F. Whiting and J.W. York, Phys. Rev. D {\bf42}, 3376 (1990).

\bibitem{York}
J.W. York, Phys. Rev. D {\bf33}, 2092 (1986).

\bibitem{ADM}
R. Arnowitt, S. Deser and C.W. Misner, {\it The Dynamics of General Relativity}, in Gravitation: An Introduction to Current Research, edited by L. Witten (Wiley, New York, 1962), pp. 227–264.

R. Arnowitt, S. Deser and C.W. Misner, {\it Republication of: The dynamics of general relativity}, General Relativity and Gravitation {\bf 40}, 1997 (2008); arXiv:gr-qc/0405109.

\bibitem{Regge}
T. Regge and C. Teitelboim, {\it Role of Surface integrals in the Hamiltonian Formulation of General Relativity}, Annals of Physics {\bf 88}, 286 (1974).

\bibitem{BrownYork}
J.D. Brown and J.W. York, {\it Quasilocal Energy and Conserved Charges Derived from the Gravitational Action}, Phys. Rev. D {\bf47}, 1407 (1993); arXiv:gr-qc/9209012.

\bibitem{WuChenLiu}
M.-F. Wu ,C.-M. Chen, J.-L. Liu and J. M. Nester, {\sl Optimal choices of reference for a quasi-local energy}, Physics Letters A {\bf 374}, 3599 (2010).

\bibitem{Lundgren} 
A.P. Lundgren, B.S. Schmekel and J.W. York, {\it Self-renormalization of the classical quasilocal energy}, Phys. Rev. D {\bf 75}, 084026 (2007).

\bibitem{rindler}
W. Rindler, {\it Relativity: Special, General, and Cosmological},  Second Edition 2006 (Oxford: Oxford University Press).

\bibitem{GibbonsHawking}
G.W. Gibbons and S.W. Hawking, {\it Cosmological Event Horizons, Thermodynamics, and Particle Creation}, Phys. Rev. D \textbf{15}, 2738 (1977).

\bibitem{Gibbons}
G.W. Gibbons and S.W. Hawking, Phys. Rev. D \textbf{15}, 2752
(1977).

\bibitem{Lanczos-Israel}
C. Lanczos, Phys. Z. 23, 539 (1922); Ann. Phys. (Leipzig) {\bf 74}, 518 (1924).

W. Israel, Nuovo Cim. B {\bf 44}, (1966) 1; Erratum-ibid B {\bf 48}, (1967) 463.

\bibitem{MisnerThorneWheelerLibro}
C.W. Misner, K.S. Thorne and J.A. Wheeler, Gravitation (Freeman, San Francisco, 1973).

%\bibitem{xin}
%M.F.Wu, C.M.Chen, J.L. Liu and J.M.Nester, Phys. Lett. A {\bf 374} (2010) 3599.





\end{thebibliography}
\end{document}